\documentclass[proceedings, preprint]{rmaa}

\usepackage{paralist}

\usepackage{psfrag,color}

\SetYear{2020}
\SetConfTitle{Robotic Observatories Workshop}

\title{Grandma: a network to coordinate them all} 

\author{S. Agayeva, 			\altaffilmark{1} 
				S. Alishov, 			\altaffilmark{1}       
				S. Antier, 				\altaffilmark{2} 
				V. R. Ayvazian,		\altaffilmark{3,4} 
				J. M. Bai, 				\altaffilmark{5}
				A. Baransky, 			\altaffilmark{6,7}
				K. Barynova, 			\altaffilmark{6,7}
				S. Basa, 					\altaffilmark{8}
				S. Beradze, 			\altaffilmark{3,4}
				E. Bertin, 				\altaffilmark{9}
				J. Berthier, 			\altaffilmark{10}
        M. Bla\v{z}ek, 	 	\altaffilmark{11} 
				M. Bo\"er,     	 	\altaffilmark{12}
				O. Burkhonov, 		\altaffilmark{13}
				A. Burrell, 			\altaffilmark{14}
				A. Cailleau, 			\altaffilmark{15}
				B. Chabert, 			\altaffilmark{14,15}
				J. C. Chen, 			\altaffilmark{16}
        N. Christensen, 	\altaffilmark{12} 
        A. Coleiro,     	\altaffilmark{2} 
        D. Corre,       	\altaffilmark{17} 
        M. W. Coughlin, 	\altaffilmark{18} 
        D. Coward,      	\altaffilmark{14} 
        H. Crisp, 				\altaffilmark{14}
        C. Delattre, 			\altaffilmark{10} 
				T. Dietrich, 			\altaffilmark{19} 
        J. G. Ducoin,   	\altaffilmark{17} 
        P. A. Duverne,  	\altaffilmark{17} 
        L. Eymar, 				\altaffilmark{12}
				P. Fock-Hang, 		\altaffilmark{10} 
        B. Gendre,      	\altaffilmark{14,27} 
        P. Hello,       	\altaffilmark{17} 
        E. J. Howell, 		\altaffilmark{14}
				R. Ya. Inasaridze,\altaffilmark{3,4}
				N. Ismailov, 			\altaffilmark{1} 
        D. A. Kann, 			\altaffilmark{11} 
        G. V. Kapanadze,	\altaffilmark{3,4}
        S. Karpov, 				\altaffilmark{20} 
        A. Klotz,       	\altaffilmark{15,21} 
        N. Kochiashvili,	\altaffilmark{3,4} 
        C. Lachaud,     	\altaffilmark{2} 
        N. Leroy,       	\altaffilmark{17} 
        A. Le Van Su, 		\altaffilmark{8}
        W. X. Li, 				\altaffilmark{22} 
				W. L. Lin, 				\altaffilmark{22} 
				P. Lognone, 			\altaffilmark{17} 
				G. Marchal-Duval, \altaffilmark{17}
        R. Marron, 				\altaffilmark{2}
				M. Ma\v{s}ek, 		\altaffilmark{20}
				J. Mo, 						\altaffilmark{22} 
				J. Moore,					\altaffilmark{14}
        D. Morris,      	\altaffilmark{23}
        R. Natsvlishvili,	\altaffilmark{3}
        K. Noysena,     	\altaffilmark{12,15} 
				N. B. Orange,    	\altaffilmark{24}
				S. Perrigault, 		\altaffilmark{10}
				A. Peyrot, 				\altaffilmark{10}
				M. Prouza, 				\altaffilmark{20}
				T. Sadibekova, 		\altaffilmark{13,25}
				D. Samadov 				\altaffilmark{1}
				A. Simon,  				\altaffilmark{26}
        C. Stachie,     	\altaffilmark{12} 
        J. P. Teng, 			\altaffilmark{10}
				P. Thierry, 			\altaffilmark{10}
        C. C. Th\"one,  	\altaffilmark{11} 
        Y. Tillayev, 			\altaffilmark{13}
        D. Turpin,      	\altaffilmark{25} 
        A. de Ugarte Postigo, \altaffilmark{11} 
				F. Vachier, 			\altaffilmark{10}
				M. Vardosanidze, 	\altaffilmark{3,4}
				V. Vasylenko, 		\altaffilmark{26}
				Z. Vidadi, 				\altaffilmark{1} 
				C. J. Wang,				\altaffilmark{5} 
        X. F. Wang,  			\altaffilmark{14} 
				S. Y. Yan, 				\altaffilmark{22} 
				J. C. Zhang, 			\altaffilmark{22} 
				J. J. Zhang, 			\altaffilmark{22} 
				And X. H. Zhang, 	\altaffilmark{22} 
}

\altaffiltext{1} {N.Tusi Shamakhy astrophysical Observatory\ Azerbaijan National Academy of Sciences, settl.Mamedaliyev, AZ 5626, Shamakhy, Azerbaijan}
\altaffiltext{2} {APC, Univ. Paris Diderot, CNRS/IN2P3, CEA/lrfu, Obs de Paris, Sorbonne Paris Cit\'e, France}
\altaffiltext{3} {E. Kharadze Abastumani Astrophysical Observatory at Ilia State University, K. Cholokashvili Av. 3/5, Tbilisi, 0162, Georgia}
\altaffiltext{4} {Samtskhe-Javakheti  State  University, Rustaveli Str. 113,  Akhaltsikhe, 0080,  Georgia}
\altaffiltext{5} {Yunnan Astronomical Observatories/Chinese Academy of Science, Kunming, 650011, China}
\altaffiltext{6} {Nuclear Physics Department\ Taras Shevchenko National University of Kyiv, 60 Volodymyrska str., Kyiv, 01601, Ukraine}
\altaffiltext{7} {Astronomical Observatory\ Taras Shevshenko National University of Kyiv, Observatorna str. 3, Kyiv, 04053, Ukraine}
\altaffiltext{8} {Laboratoire d'Astrophysique de Marseille, UMR 7326, CNRS, Universit\'e d'Aix Marseille, 38, rue Fr\'ed\'eric Joliot-Curie,Marseille, France}
\altaffiltext{9} {UPMC-CNRS, UMR7095, Institut d'Astrophysique de Paris, 75014 Paris, France}
\altaffiltext{10}{AGORA observatoire des Makes, AGORA, 18 Rue Georges Bizet, Observatoire des Makes, 97421 La Rivi\`ere, France}
\altaffiltext{11}{Instituto de Astrof\'isica de Andaluc\'ia (IAA-CSIC), Glorieta de la Astronom\'ia s/n, 18008 Granada, Spain}
\altaffiltext{12}{ARTEMIS UMR 7250 UCA CNRS OCA, boulevard de l'Observatoire, CS 34229, 06304 Nice CEDEX 04, France}
\altaffiltext{13}{Ulugh Beg Astronomical Institute, Uzbekistan Academy of Sciences, Astronomy str. 33, Tashkent 100052, Uzbekistan}
\altaffiltext{14}{OzGrav-UWA, University of Western Australia, School of Physics, M013, 35 Stirling Highway, Crawley, WA 6009, Australia}
\altaffiltext{15}{IRAP, Universit\'e de Toulouse, CNRS, UPS, 14 Avenue Edouard Belin, F-31400 Toulouse, France}
\altaffiltext{16}{WuZhou University, WuZhou, GuangXi, 543000, China}
\altaffiltext{17}{IJCLab, Univ Paris-Saclay, CNRS/IN2P3, Orsay, France}
\altaffiltext{18}{School of Physics and Astronomy, University of Minnesota, Minneapolis, Minnesota 55455, USA}
\altaffiltext{19}{Institut f\"{u}r Physik und Astronomie, Universit\"{a}t Potsdam, Haus 28, Karl-Liebknecht-Str. 24/25, 14476, Potsdam, Germany}
\altaffiltext{20}{FZU – Institute of Physics of the Czech Academy of Sciences, Prague, Czech Republic}
\altaffiltext{21}{Universit\'e Paul Sabatier Toulouse III, Universit\'e de Toulouse, 118 route de Narbonne, 31400 Toulouse, France}
\altaffiltext{22}{Physics Department and Astronomy Department, Tsinghua University, Beijing, 100084, China}
\altaffiltext{23}{University of the Virgin Islands, 2 Brewer's Bay, 00902 St Thomas, US VI}
\altaffiltext{24}{OrangeWave Innovative Science, LLC, Moncks Corner, SC 29461, USA}
\altaffiltext{25}{AIM, CEA, CNRS, Universit\'e Paris-Saclay, Universit\'e Paris Diderot, Sorbonne Paris Cit\'e, F-91191 Gif-sur-Yvette, France}
\altaffiltext{26}{Astronomy and Space Physics Department\ Taras Shevchenko NationalUniversity of Kyiv, 60 Volodymyrska str., Kyiv, 01601, Ukraine}
\altaffiltext{27}{Corresponding author. Bruce.Gendre@uwa.edu.au}

\shortauthor{Agayeva et al.}
\shorttitle{The GRANDMA collaboration}

\listofauthors{S. Agayeva, S. Alishov, S. Antier, V. R. Ayvazian, J. M. Bai, A. Baransky, K. Barynova, S. Basa, S. Beradze, E. Bertin, J. Berthier, M. Bla\v{z}ek, M. Bo\"er, O. Burkhonov, A. Burrell, A. Cailleau, B. Chabert, J. C. Chen, N. Christensen, A. Coleiro, D. Corre, M. W. Coughlin, D. Coward, H. Crisp, C. Delattre, T. Dietrich, J. G. Ducoin, P. A. Duverne, L. Eymar, P. Fock-Hang, B. Gendre, P. Hello, E. J. Howell, R. Ya. Inasaridze, N. Ismailov, D. A. Kann, G. V. Kapanadze, S. Karpov, A. Klotz, N. Kochiashvili, C. Lachaud, N. Leroy, A. Le Van Su, W. X. Li, W. L. Lin, P. Lognone, G. Marchal-Duval, R. Marron, M. Ma\v{s}ek, J. Mo, J. Moore, D. Morris, R. Natsvlishvili, K. Noysena, N. B. Orange, S. Perrigault, A. Peyrot, M. Prouza, T. Sadibekova, D. Samadov, A. Simon, C. Stachie, J. P. Teng, P. Thierry, C. C. Th\"one, Y. Tillayev, D. Turpin, A. de Ugarte Postigo, F. Vachier, M. Vardosanidze, V. Vasylenko, Z. Vidadi, C. J. Wang, X. F. Wang, S. Y. Yan, J. C. Zhang, J. J. Zhang, \& X. H. Zhang}

\indexauthor{Agayeva, S.}
\indexauthor{Alishov, S.}
\indexauthor{Antier, S.}
\indexauthor{Ayvazian, V. R.}
\indexauthor{Bai, J. M.}
\indexauthor{Baransky, A.}
\indexauthor{Barynova, K.}
\indexauthor{Basa, S.}
\indexauthor{Beradze, S.}
\indexauthor{Bertin, E.}
\indexauthor{Berthier, J.}
\indexauthor{Bla\v{z}ek, M.}
\indexauthor{Bo\"er, M.}
\indexauthor{Burkhonov, O.}
\indexauthor{Burrell, A.}
\indexauthor{Cailleau, A.}
\indexauthor{Chabert, B.}
\indexauthor{Chen, J. C.}
\indexauthor{Christensen, N.}
\indexauthor{Coleiro, A.}
\indexauthor{Corre, D.}
\indexauthor{Coughlin, M. W.}
\indexauthor{Coward, D.}
\indexauthor{Crisp, H.}
\indexauthor{Delattre, C.}
\indexauthor{Dietrich, T.}
\indexauthor{Ducoin, J. G.}
\indexauthor{Duverne, P. A.}
\indexauthor{Eymar, L.}
\indexauthor{Fock-Hang, P.}
\indexauthor{Gendre, B.}
\indexauthor{Hello, P.}
\indexauthor{Howell, E. J.}
\indexauthor{Inasaridze, R. Ya.}
\indexauthor{Ismailov, N.}
\indexauthor{Kann, D. A.}
\indexauthor{Kapanadze, G. V.}
\indexauthor{Karpov, S.}
\indexauthor{Klotz, A.}
\indexauthor{Kochiashvili, N.}
\indexauthor{Lachaud, C.}
\indexauthor{Leroy, N.}
\indexauthor{Le Van Su, A.}
\indexauthor{Li, W. X.}
\indexauthor{Lin, W. L.}
\indexauthor{Lognone, P.}
\indexauthor{Marchal-Duval, G.}
\indexauthor{Marron, R.}
\indexauthor{Ma\v{s}ek, M.}
\indexauthor{Mo, J.}
\indexauthor{Moore, J.}
\indexauthor{Morris, D.}
\indexauthor{Natsvlishvili, R.}
\indexauthor{Noysena, K.}
\indexauthor{Orange, N. B.}
\indexauthor{Perrigault, S.}
\indexauthor{Peyrot, A.}
\indexauthor{Prouza, M.}
\indexauthor{Sadibekova, T.}
\indexauthor{Samadov, D.}
\indexauthor{Simon, A.}
\indexauthor{Stachie, C.}
\indexauthor{Teng, J. P.}
\indexauthor{Thierry, P.}
\indexauthor{Th\"one, C. C.}
\indexauthor{Tillayev, Y.}
\indexauthor{Turpin, D.}
\indexauthor{de Ugarte Postigo, A.}
\indexauthor{Vachier, F.}
\indexauthor{Vardosanidze, M.}
\indexauthor{Vasylenko, V.}
\indexauthor{Vidadi, Z.}
\indexauthor{Wang, C. J.}
\indexauthor{Wang, X. F.}
\indexauthor{Yan, S. Y.}
\indexauthor{Zhang, J. C.}
\indexauthor{Zhang, J. J.}
\indexauthor{Zhang, X. H.}

\abstract{GRANDMA is  an international project that coordinates telescope observations of transient sources with large localization uncertainties. Such sources include gravitational wave events, gamma-ray bursts and neutrino events. GRANDMA currently coordinates 25 telescopes (70 scientists), with the aim of optimizing the imaging strategy to maximize the probability of identifying an optical counterpart of a transient source. This paper  describes the motivation for the project, organizational structure, methodology and initial results.}

\resumen{GRANDMA es un proyecto internacional que coordina las observaciones de telescopio de fuentes transitorias con grandes
incertidumbres de localización. Dichas fuentes incluyen eventos de ondas gravitacionales, explosiones de rayos gamma y neutrinos.
GRANDMA actualmente coordina 25 telescopios (70 científicos), con el objetivo de optimizar las imágenes. La
estrategia es maximizar la probabilidad de identificar una contraparte óptica de una fuente transitoria. Este art\'{\i}culo
describe la motivación del proyecto, la estructura organizacional, la metodología y los resultados iniciales.}

\addkeyword{Gravitational-waves: general}
\addkeyword{Robotic Observatories}

\begin{document}
\maketitle

\section{Introduction}

Astrophysics is usually considered as the science of the extremes. We, astronomers, are studying the most massive clusters of matters, the strongest fields, the hottest places, the coldest areas, the content of the emptiest volumes. We perform all of that by doing observations, which are usually impossible to reproduce in laboratories. Among the classification of observations, there is a key discrepancy: transient astronomy versus secular astronomy. While the latter one allows for repeated observations and can accommodate a few misses, the former is far more challenging and needs a complete new approach of the astronomical observation practices. Firstly, the transient phenomena are usually visible in the sky for only less than a second up to days, before vanishing into the darkness of the Universe. Each missed observation is a hole in the data flow, and possibly the reason why an event cannot be understood or a discovery not made. Gathering most of the data from any transient event is a top priority goal of any collaboration or astronomer working in this research domain.

If one is focusing on gravitational wave events only, the situation is even more complicated, as the position of the sources is poorly known. The localization of the first gravitational wave (GW) observation of a coalescing binary of neutron stars (BNS), GW170817, by the Advanced LIGO~\citep{TheLIGOScientific:2014jea} and Advanced Virgo~\citep[aLIGO/Virgo;][]{TheVirgo:2014hva} detectors, which was to within 28\,deg$^2$ \citep{LSC_BNS_2017PhRvL}, was very misleading in that sense. It was the prompt joint observations by \textit{Fermi}-GBM and \textit{INTEGRAL} of the short GRB 170817A \citep{LSC_GW_GRB_2017ApJ,goldstein_ordinary_2017,savchenko_integral_2017} which allowed such a precise position and the following observations. In most of the cases, gravitational wave events, currently detected by the LIGO-Virgo Collaboration, have typical error boxes that cover hundreds of square degrees in the sky. For the telescopes having a field of view smaller than 1 deg$^2$, the follow-up of such poorly localized events is a very complex task. In a general manner, the transient astronomy has to make use of coordinated and smart observational strategies combining both fast and wide-field angle automated facilities with more sensitive narrow-field of view facilities in order to obtain the best scientific data set. This, however can lead to very valuable and ground-breaking observations \citep[see e.g.][]{LSC_MM_2017ApJ,Andreoni_2017PASA,Hallinan_2017Sci,Kasliwal_2017Sci,Troja_2017Natur,JuBa2015,KaKa2019}. 

In order to solve these technical issues, we initiated the Global Rapid Advanced Network Devoted to the Multi-messenger Addicts \citep[GRANDMA;][]{ant20} in late 2018. The purpose of this article is to present the collaboration, its organization, and the results obtained so far. The paper is organized accordingly, with these three topics presented in Sections \ref{collab}, \ref{orga}, and \ref{results} respectively. A small conclusion presents the improvements we identified and what we are trying to address in the next months.

\section{GRANDMA consortium}
\label{collab}

\subsection{A network of telescope}

\begin{figure*}[!t]
  \includegraphics[width=16cm]{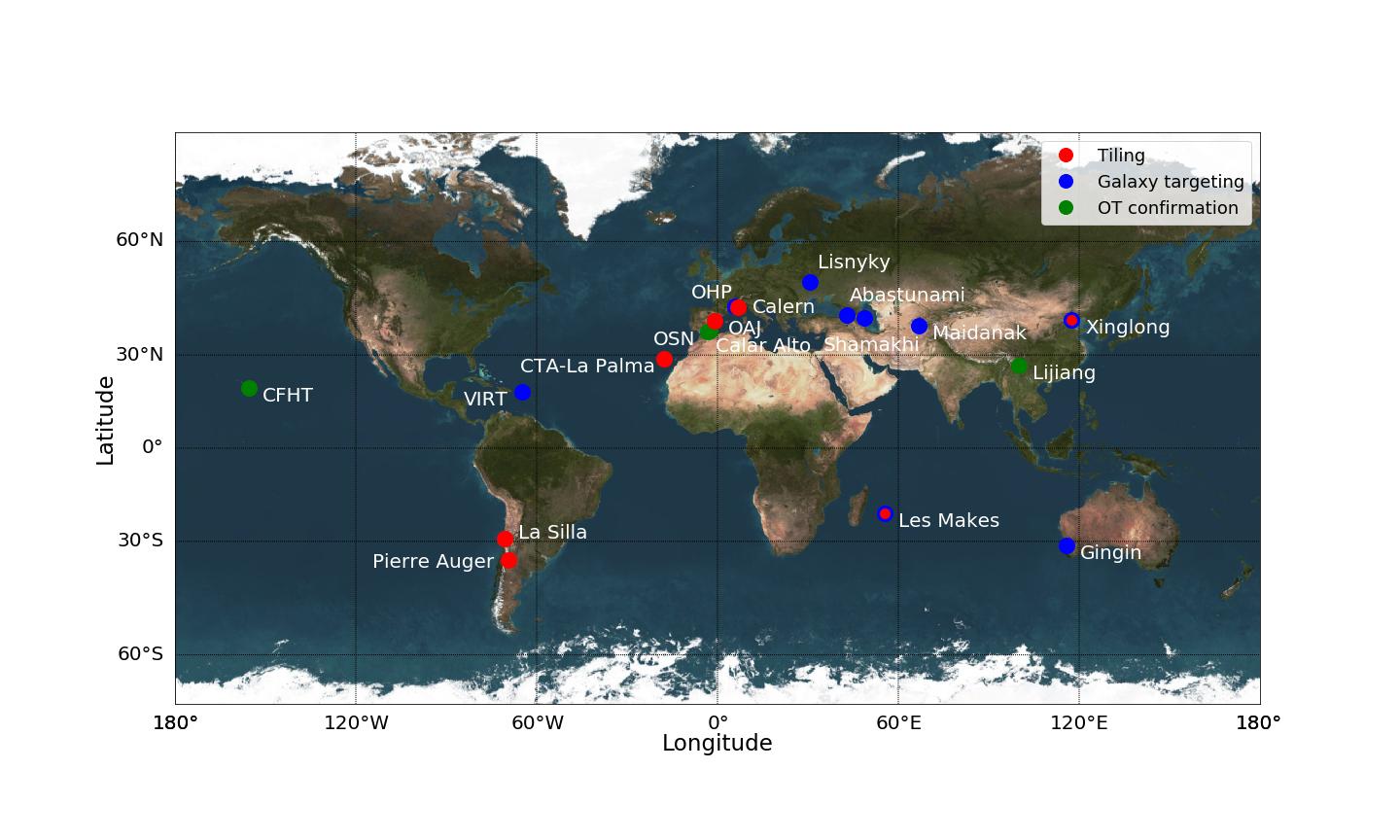}
  \caption{\label{fig1} The wold-wide network of the GRANDMA collaboration.}
  
\end{figure*}

The GRANDMA consortium is a world-wide network of 25 telescopes with both photometric and spectroscopic facilities (see Figure \ref{fig1}), with a large amount of time allocated for observing transient alerts as a telescope network (see Table~\ref{tab:GRANDMAtelphoto}). It includes 17 observatories, 26 institutions and groups from nine countries. Most of the allocated time available to GRANDMA is proprietary, i.e. warranted recurring time. Because of that, the collaboration can cover thousands of square degrees within 24 hours.

Many telescopes of our network were built in the 1960's and 1970's, and at that time the largest telescopes at an observatory generally still had a human sized mirror: it was very common to have one-meter (or smaller) class instruments at each observatory, with several observatories in each country. These ``small'' aperture telescopes are used for detection of mildly bright ($\sim$\,20-22\,mag) objects. Despite their age, they are still operating, either manually or robotically. The GRANDMA network also has access to four wide-field telescopes (field of view [FoV] $>\mathrm{1\,deg}^2$) located on three continents, with the TAROT network \citep{2008PASP..120.1298K,2018cosp...42E2475N} and the OAJ-T80 telescope. 

The two groups of telescopes are used to rapidly find counterpart candidates, either tiling the localization area of the GW alerts, or targeting host-galaxy candidates. Small field of view telescopes are also used to confirm an association and to perform dedicated follow-up observations, either photometrically or spectroscopically. We indeed have access to spectrometers. In the case of a confident detection of an electromagnetic counterpart to a GW event, GRANDMA has access to four 2m-class telescopes in three different time zones with spectroscopic capabilities, and has been granted several observation hours on larger instruments such as the CFHT or the GTC for the duration of the Third Acquisition Run of the LIGO-Virgo Collaboration (hereafter O3 run). Because of our very conservative approach on observation requests, and contrary to most teams willing to participate to O3, we managed to keep a fair amount of available time for a last minute interesting event. In 24 hours and for a typical night, thanks to an extensive distribution on Earth the GRANDMA network is able to access more than 72\% of sky to a limiting magnitude of 18 mag.

\begin{table*}
\begin{tabular}{ccccc}
Telescope      & Location        & Aperture & FOV             & $3-\sigma$ limit   \\
Name           &                 & (m)      &  (deg)          & (AB mag)           \\
\hline
\hline
FRAM/Auger     & Pierre Auger Obs.  & 0.30 & $1.0\times1.0$ & 18.0 in 60s (R$_C$) \\
TAROT/TCH      & La Silla Obs.      & 0.25 & $1.85\times1.85$ & 18.0 in 60s (Clear) \\
VIRT           & Etelman Obs.       & 0.50 & $0.27\times0.27$ & 19.0 in 120s (Clear)\\
CFHT/WIRCAM    &    CFH Obs.        & 3.6  & $0.35\times0.35$ & 22.0 in 200s (J)    \\
CFHT/MEGACAM   &     CFH Obs.       & 3.6  & $1.0\times1.0$   & 23.0 in 200s (r$^\prime$ )\\
Zadko          & Gingin Obs.        & 1.00 & $0.17\times0.12$ & 20.5 in 40s (Clear) \\
TNT            & Xinglong Obs.      & 0.80 & $0.19\times0.19$ & 19.0 in 300s (R$_C$)\\
Xinglong-2.16  & Xinglong Obs.      & 2.16 & $0.15\times0.15$ & 21.0 in 100s (R$_C$)\\
GMG-2.4        & Lijiang Obs.       &  2.4 & $0.17\times0.17$ & 22.0 in 100s (R$_C$)\\
UBAI/NT-60     & Maidanak Obs.      & 0.60 & $0.18\times0.18$ & 18.0 in 180s (R$_C$)\\
UBAI/ST-60     & Maidanak Obs.      & 0.60 & $0.11\times0.11$ & 18.0 in 180s (R$_C$)\\
TAROT/TRE      & La Reunion         & 0.18 & $4.2\times4.2$   & 16.0 in 60s (Clear) \\
Les Makes/T60  & La Reunion.        & 0.60 & $0.3\times0.3$   & 19.0 in 180s (R$_C$)\\
Abastumani/T70 & Abastumani Obs.    & 0.70 & $0.5\times0.5$   & 18.2 in 60s (R$_C$) \\
Abastumani/T48 & Abastumani Obs.    & 0.48 & $0.33\times0.33$ & 15.0 in 60s (R$_C$) \\
ShAO/T60       & Shamakhy Obs.      & 0.60 & $0.28\times0.28$ & 19.0 in 300s (R$_C$)\\
Lisnyky/AZT-8  & Kyiv Obs.          & 0.70 & $0.38\times0.38$ & 20.0 in 300s (R$_C$) \\
TAROT/TCA      & Calern Obs.        & 0.25 & $1.85\times1.85$ & 18.0 in 60s (Clear) \\
IRIS           & OHP                & 0.5  & $0.4\times0.4$   & 18.5 in 60s (r$^\prime$)\\
T120           & OHP                & 1.20 & $0.3\times0.3$   & 20.0 in 60s (R)     \\
OAJ/T80        & Javalambre Obs.    & 0.80 & $1.4\times1.4$   & 21.0 in 180s (r$^\prime$)\\
OSN/T150       & Sierra Nevada Obs. & 1.50 & $0.30\times0.22$ & 21.5 in 180s (R$_C$)\\
CAHA/2.2m      & Calar Alto Obs.    & 2.20 & $0.27^\circ$     & 23.7 in 100s (r$^\prime$)\\
FRAM/CTA       & La Palma Obs.      & 0.25 & $0.5\times0.5$   & 17.0 in 90s (R$_C$) \\
\hline
\end{tabular}
\caption{List of telescopes of the GRANDMA consortium and their photometric performance when using their standard setup. Updated from Antier et al.(2020).}
\label{tab:GRANDMAtelphoto}
\end{table*}

\subsection{GRANDMA infrastructure: ICARE in a nutshell}

To support the GRANDMA operations and make the collaboration competitive in multi-messenger time-domain astronomy, we developed ICARE (``Interface and Communication for Addicts of the Rapid follow-up in multi-messenger Era''), an e-infrastructure composed by a set of tools and an interface to help coordinating telescope observations, easing the data reduction and analysis, and allowing some automated communication with the transient community. ICARE is also a heritage from previous work done (e.g TAROT, Bourez-Laas et al. 2008, GROWTH and Pan-STARRS, Kasliwal et al. 2019a) and can be adapted to any future missions \citep[as the SVOM mission;][]{wei16}. ICARE has been continuously improved during the third observational campaign organized by LVC and can be now extensively used by any consortium and group intending to monitor facilities for rapid follow-up of multi-messenger alerts.  

The central system is responsible for ingesting the transient alerts coming from various platforms, such as GCN notices. The system can also receive the current status of the various observatories in terms of availability and weather forecast. Cross-matching with pre-established programs of the various teams (in terms of acceptable Target of Opportunity rate, and allocation time), ICARE distributes a dedicated observation plan to all the telescopes within the network via a broker delivering standardized VO events. At the same time scale, the information from the alert and the pre-programmed observations are stored into the database. This information can be visualized on a dedicated web portal. The data analysis is at the charge of the different telescope teams but a generic pipeline is under development to harmonize the processes (this is outside the scope of this paper). Sub-images containing interesting sources and photometric measurements are reported in the database as well as the list of observations that have been performed. In case of the detection of interesting transients (kilonova, electromagnetic counterparts of neutrino sources, ...), the scientists on duty can send a command to specific telescopes to ask for further observations of the source via the web portal.

\subsubsection{Communications}

We have decided to rely on the recommendations of the International Virtual Observatory Alliance (IVOA, https://wiki.ivoa.net/), and to use their protocols for communications. In particular, we have selected the {\it VO} protocol\footnote{https://wiki.ivoa.net/twiki/bin/view/ IVOA/VOEventTwoPointZero}, as it implies standardized, machine-readable messages containing observational properties of an astrophysical source. This goes towards a better standardization of the communications at the professional level between observatories, one of the core missions of the IVOA, and is followed by a growing number of major actors in the transient fields \citep[e.g. LIGO-Virgo, FRB community;][]{2017arXiv171008155P}. 

The {\it VO} standard defines the structure of the XML packet, but leaves free the choice of additional keywords inside the message (the data model, using the IVOA terminology). For ICARE, we created a new data model including a few mandatory extra keywords to fulfill our needs, and we are sending all our messages using it. None are expected to be acknowledged, but it is assumed that these messages are listened, parsed, and processed in real time.

Because of the heterogeneity of the telescope infrastructures, we have decided to rely on the most flexible possible technology, the http protocol, using web pages. An automated instrument will be able to simulate a 'post' request with a list of specified keywords, while an human user will use the interface provided by the web pages. ICARE requests the following information: the status of the observatory, the accepted and refused requests, the observations done, and in case of local processing, any transient detected. It is to the choice of the user to decide the frequency of reports, however a good coordination implies a frequent update of each part of the information.

\subsubsection{Classes of users}
One of the main purpose of the web interface is to operate the telescopes from the various pages of the interface. In order to prevent multiple users from sending conflicting requests, we defined three classes of users of the interface.

\begin{itemize}
\item The general users. These are standard users, who can follow the operations and read the information, but cannot operate the telescopes. If they want to send data or results, they need to use the dedicated pages, and can only create new information, neither alter or delete it.
\item The supervisor. This is a super-user, who has all the rights of a general user, and can also operate the telescopes. ICARE assumes there are shifts among users, and that only one supervisor at a time exists. There is however a possibility to request an upgrade of status from the web interface in case of emergency. The supervisor can alter information if needed considering the last results obtained. Within the collaboration, the supervisors are called Follow-up Advocates (FA).
\item The administrators. There are a set of administrator pages aimed at handling the content of the database (such as the list of co-Is, the instruments involved), and fixing some issues into the database without the need to physically log into the server. Technically, they have the maximal power on the interface, as they can remove some information if needed.
\end{itemize}

\subsubsection{Web interface and human monitoring}

ICARE offers a graphical user interface to allow for human supervision at any time. Technically, the interface is built with web 2.0 technologies, mainly PHP7, HTML5/CSS3 and jQuery. In practice, we used HTML5 for providing the general layout, CSS for the web pages visual style and appearance, JavaScript (mainly through jQuery) functions to make dynamic web pages, and PHP7 to communicate with the database. Besides the main page, which provides some public elements about the collaboration, each page is protected by a login scheme to prevent unauthorized access. Once the user is recognized, the full web interface is accessible without further limitation. ICARE is accessible from any web browser and has been tested under Firefox, Chrome and Edge. It can also be accessed via smartphones and tablets. 

The interface is organized by tabs, reflecting the various steps of a standard follow-up campaign: the observation of the whole error box with various instruments; the detections and reporting of candidates; the study of a dedicated counterpart; and the reporting tools (such as a GCN parser). Several plug-ins ease the visualization of the information.

The interface reacts to the class of the user, enabling or disabling functions and pages accordingly. The system is highly reactive, but there are ways, if needed, to overcome this by becoming FA on demand.

\section{Observation strategy}
\label{orga}

\subsection{Observation strategy for GW alerts}

As explained earlier, the GRANDMA observation strategy uses different approaches depending on the FoV of the telescopes. The core of the observation plans relies on \texttt{gwemopt} \citep{CoTo2018}, an open-source software designed to optimize the scheduling of target-of-opportunity observations using either synoptic or galaxy-targeted based searches.
Briefly, for synoptic survey instruments, it divides the skymap into a grid of ``tiles'' the size and shape of the FoV of each telescope.
For each tile, the available segments for observation are computed, including accounting for their rising and setting, in addition to their distance from the moon.
Observations are scheduled by computing a weight based on the three-dimensional probability distribution from the GW alert \citep{Singer:2016eax}.
For narrow-FoV telescopes with nearby alerts ($\lesssim300$ Mpc), galaxy-targeted observations are performed, where one galaxy at a time is imaged within the three-dimensional probability distribution as suggested by \citet{2013ApJ...767..124N,2016ApJ...820..136G}. Each observation is processed by the reduction pipelines to find transient candidates. Each team of the GRANDMA consortium can use their own detection pipeline; a subset have low-latency reduction processes (the TAROT network as an example, \citealt{2018cosp...42E2475N}). Each telescope team is supposed to report calibrated magnitudes of the counterpart candidates into the database using the dedicated webpages. Each of these candidates is then evaluated and if needed other observations can be requested, as follow-up, to classify the object.

\subsection{Organization of the follow-up procedure}

Various organizations are possible for the Follow-up Advocates (FAs): they can go on an event by event basis, or on a time to time basis. We chose the second option, and our FAs are shifters, working for one week inside a team of 4. In case of an alert during their shift, the first responsibility of the FA is to check that the observation plan has been computed, sent to the network of observatories and has been well-received by them. The FA has then to follow all notices (GCNs) related to the alerts currently in hand. If an optical transient is found during the campaign (by GRANDMA or by others) and the first observations can be consistent with a kilonova or GRB afterglow signal, the FA may ask to follow it up by dedicated telescopes of the network. This may require the interruption of current observations for some telescopes and the modification of the observation plan. Lastly, the FA is in charge of reporting everything that happened during their shift, either internally through a system of logbooks, or toward the whole community using an automatic writing tool generating the GCNs related to GRANDMA observations and possible discoveries.

\section{GRANDMA electromagnetic follow-up campaign of O3 run A}
\label{results}

The third aLIGO/Virgo observing run (O3) started in April 2019, and is still ongoing at the time of writing, after a one-month pause in October 2019. The GRANDMA collaboration followed up 27/33 alerts during the first six months of the O3 campaign based on LIGO-Virgo low latency results (see the LIGO-Virgo user guide~\footnote{https://emfollow.docs.ligo.org/userguide/} for more information). 

From April to September 2019, the delay between the GW trigger time and the first observation by the TAROT network decreased from more of twelve hours to $15-30$ minutes. The most rapid follow-up was for S190515ak with a delay of 16 minutes. The coverage of GRANDMA also improved during the same interval, going from 100 deg$^2$ to 300 deg$^2$ covered to 540 deg$^2$ in 68~hours thanks to a more efficient coordination of the telescopes.

Despite all our efforts, no significant counterparts were found by the GRANDMA consortium for any alert (in agreement with observations by other groups). We can however explain that with two simple arguments. We first note that none of our follow-ups have led to a complete coverage of any error box. This can be explained by the fact that Sun and Moon constraints do not allow us to perform full coverage (see the example of S190924h). In addition, the luminosity distances of the alerts sent by LIGO-Virgo are mostly above 150 Mpc (the closest binary neutron star candidate was estimated to lie at $156\pm41$ Mpc), where the galaxy catalog we are using to perform galactic tiling is no longer complete.

\section{Discussion: future improvements}
\label{future}

There has been a significant effort to optimize single-telescope observations (for example \citealt{2016ApJ...820..136G,CoTo2018,2017ApJ...838..108R,2016A&A...592A..82G}) for large target-of-opportunity sky areas in the context of time-domain astronomy. We have experienced that the situation is far more complex in case of a network of telescopes. Several parts can be strongly improved.

Firstly, the differences in sensitivity and FoV potentially lead to significantly different search strategies. We are working toward a globalized detection pipeline, and a precise cross calibrations of the instruments to reduce these effects. On a more distant scale, a future milestone is the integration of an automatic evaluation of the nature of optical transients with spectroscopic tools and results of rapid evaluation of photometric light-curves with machine learning \citep{2019arXiv190400014M}. 

Secondly, a network doing both synoptic and galaxy-targeted observations implies the necessity to decide if covering overlapping parts of the sky is appropriate or not. A centralized scheduler supervising all the telescopes will ideally be the best option to solve this issue. But this is at the cost of having to significantly modify the robotic system used by the telescopes. It is also very important to have the larger completeness of the catalog we are using, and one solution to increase the current one would be to use the MANGROVE catalog \citep{duc20}

A last axis of development is the extension of the interface to other triggering instruments, such as the Fermi-GBM, as well as the extension of the web portal for neutrino alerts.

 Another work can consist of combining independent detections from GRANDMA or other optical surveys and gamma-ray surveys with the alerts and evaluate compatibility of time delays, localization and light curve profiles.

\section{Conclusion}

We have presented the GRANDMA consortium. It consists of a global network of 25 telescopes with both photometric and spectroscopic facilities, located in 17 different observatories, and aims to face the challenges of time-domain astronomy. Our main scientific goal is currently the detection of gravitational wave counterparts inside very large error boxes as presented in \citet{Networkpaper}. The network and the large availability of the telescopes allow for a 24-hour coverage, which helps to rapidly scan the GW sky localization area to $\sim18$th magnitude. The tools have been set up for monitoring the full network, including the joint agreement between the teams, the web interface, the uniform distribution of the observation plan, and the 24~hr follow-up advocates on duty to revise observation plans in case of interesting candidates and updated sky localization areas.

\section*{Acknowledgments}
Parts of this research were conducted by the Australian Research Council Centre of Excellence for Gravitational Wave Discovery (OzGrav), through project number CE170100004. AdUP and CCT acknowledge support from Ram\'on y Cajal fellowships RyC-2012-09975 and RyC-2012-09984 and the Spanish Ministry of Economy and Competitiveness through project AYA2017-89384-P. DAK acknowledges support from the Spanish research project AYA2017-89384-P and RTI2018-098104-J-I00 (GRBPhot). DM, and NBO acknowledge support from NASA-MUREP-MIRO grant NNX15AP95A, and NSF EiR AST Award 1901296. DT is supported by the CNES Postdoctoral Fellowship at Laboratoire AIM/CEA-Saclay. MB acknowledges funding as ``personal tecnico de apoyo'' under fellowship number PTA2016-13192-I. MC is supported by the David and Ellen Lee Postdoctoral Fellowship at the California Institute of Technology. MP, SK and MM are supported by European Structural and Investment Fund and the Czech Ministry of Education, Youth and Sports (Projects CZ.02.1.01/0.0/0.0/16\_013/0001402, CZ.02.1.01/0.0/0.0/16\_013/0001403 and CZ.02.1.01/0.0/0.0/15\_003/0000437). RYaI, VRA and GVK are supported by Shota Rustaveli National Science Foundation of Georgia, grant No RF-18-1193. SA is supported by the CNES Postdoctoral Fellowship at Laboratoire AstroParticule et Cosmologie. SA, AC, CL and RM acknowledge the financial support of the UnivEarthS Labex program at Sorbonne Paris Cit\'e (ANR-10-LABX-0023 and ANR-11-IDEX-0005-02). SA and NL acknowledge the financial support of the Programme National Hautes Energies (PNHE). IRiS has been carried out thanks to the support of the OCEVU Labex (ANR-11-LABX-0060) and the A*MIDEX project (ANR-11-IDEX-0001-02) funded by the ``Investissements d'Avenir'' French government program. IRiS and T120 thank all the Observatoire de Haute-Provence staff for the permanent support. TAROT has been built with the support of the Institut National des Sciences de l'Univers, CNRS, France. TAROT is funded by the CNES and thanks the help of the technical staff of the Observatoire de Haute Provence, OSU-Pytheas.


\begin{thebibliography}
\bibitem[Aasi et al.(2015)]{TheLIGOScientific:2014jea} Aasi, J., et al. 2015, Class. Quant. Grav., 32, 074001
\bibitem[Abbott et al.(2017a)]{LSC_MM_2017ApJ} Abbott, B.~P., Abbott, R., Abbott, T.~D. et al. 2017a, \apj, 848, L12
\bibitem[Abbott et al.(2017b)]{LSC_GW_GRB_2017ApJ} Abbott, B.~P., Abbott, R., Abbott, T.~D. et al. 2017b, \apj, 848, L13
\bibitem[Abbott et al.(2017c)]{LSC_BNS_2017PhRvL} Abbott, B.~P., Abbott, R., Abbott, T.~D. et al. 2017c, \prl, 119, 161101
\bibitem[Acernese et al.(2015)]{TheVirgo:2014hva} Acernese, F., et al., 2015, Class. Quant. Grav., 32, 024001
\bibitem[Andreoni et al.(2017)]{Andreoni_2017PASA} Andreoni, I., Ackley, K., Cooke, J., et al., 2017, PASA, 34, e069
\bibitem[Antier et al.(2020)]{ant20} Antier, S., Agayeva, S., Aivazyan, V. et al. 2020, MNRAS, in press
\bibitem[Bourez-Laas et al.(2008)]{bou08} Bourez-Laas, M., Vachier, F., Klotz, A., Damerdji, Y., \& Bo{\"e}r, M., 2008, Proceedings of the SPIE, 7019, 701918
\bibitem[Coughlin et al.(2018)]{CoTo2018} Coughlin, M.W., Tao, D., Chan, M.L., et al. 2018, MNRAS, 478, 692
\bibitem[Coughlin et al.(2019)]{Networkpaper} Coughlin, M.W., Antier, S., Corre, D. et al. 2019, MNRAS, 489, 5775
\bibitem[Ducoin et al.(2020)]{duc20} Ducoin, J.-G., Corre, D., Leroy, N., \& Le Floch, E. 2020, MNRAS, in press
\bibitem[Gehrels et al.(2016)]{2016ApJ...820..136G} Gehrels, N., Cannizzo, J.K., Kanner, J. et al. 2016, \apj, 820, 136
\bibitem[Ghosh et al.(2016)]{2016A&A...592A..82G} Ghosh, S., Bloemen, S., Nelemans, G. et al. 2016, \aap; 592, A82
\bibitem[Goldstein et al.(2017)]{goldstein_ordinary_2017} Goldstein, A., Veres, P., Burns, E., et al. 2017, \apj, 848, L14
\bibitem[Hallinan et al.(2017)]{Hallinan_2017Sci} Hallinan, G., Corsi, A., Mooley, K.~P., et al. 2017, Science, 358, 1579
\bibitem[Just et al.(2015)]{JuBa2015} Just, O., Bauswein, A., Pulpillo, R., et al. 2015, MNRAS, 448, 541
\bibitem[Kasliwal et al.(2017)]{Kasliwal_2017Sci} Kasliwal, M.~M., Nakar, E., Singer, L.~P., et al. 2017, Science, 358, 1559
\bibitem[Kasliwal et al.(2019a)]{growth} Kasliwal, M. M., Cannella, C., Bagdasaryan, A., et al. 2019, PASP, 131, 38003
\bibitem[Kasliwal et al.(2019b)]{KaKa2019} Kasliwal, M., Kasen, D., Lau, R., et al. 2019, MNRAS, submitted
\bibitem[Klotz et al.(2008)]{2008PASP..120.1298K} Klotz, A., Bo{\"e}r, M., Eysseric, J., et al., 2008, PASP, 120, 874
\bibitem[Muthukrishra et al.(2019)]{2019arXiv190400014M} Muthukrishna, D., Narayan, G., Mandel, K.S. et al. 2019, PASP, 131, 118002
\bibitem[Nissanke et al.(2013)]{2013ApJ...767..124N} Nissanke, S., Kasliwal, M., \& Georgieva, A., 2013, \apj, 767, 124
\bibitem[Noysena et al.(2019)]{2018cosp...42E2475N} Noysena, K., Klotz, A., Boer, M., et al. 2019, 42nd COSPAR Scientific Assembly. Held 14-22 July 2018, in Pasadena, California, USA, arxiv:1910.02770
\bibitem[Petroff et al.(2017)]{2017arXiv171008155P} Petroff, E., Houben, L., Bannister, K., et al. 2017, Technical document about VOEvent, arxiv:1710.08155
\bibitem[Rana et al.(2017)]{2017ApJ...838..108R} Rana, J., Singhal, A., Gadre, B. et al. 2017, \apj, 838, 108
\bibitem[Savchenko et al.(2017)]{savchenko_integral_2017} Savchenko, V., Ferrigno, C., Kuulkers, E., et al. 2017, \apj, 848, L15
\bibitem[Singer et al.(2016)]{Singer:2016eax} Singer, L.P. et al. 2016, \apj, 829, L15
\bibitem[Troja et al.(2017)]{Troja_2017Natur} Troja, E., Piro, L.,  van Eerten, H., et al. 2017, Nature, 551, 71
\bibitem[Wei et al.(2016)]{wei16} Wei, J., Cordier, B., Antier, S. et al. 2016, Proceedings of the Workshop held from 11th to 15th April 2016 at Les Houches School of Physics, France, arxiv:1610.06892
\end{thebibliography}
\end{document}